\documentclass[conference]{IEEEtran}

\ifCLASSINFOpdf
   \usepackage[pdftex]{graphicx}
\else
\fi

\begin{document}
%
\title{Charge transport and current fluctuations in bacteriorhodopsin based nanodevices}

\author{\IEEEauthorblockN{J.-F. Millithaler, E. Alfinito and L. Reggiani}
\IEEEauthorblockA{Dipartimento di Ingegneria dell'Innovazione, Universit\`a del Salento,
Lecce, Italy\\
CNISM- Consorzio Nazionale Interuniversitario per le Scienze Fisiche della Materia, Roma, Italy\\
Email : jf.millithaler@unisalento.it, eleonora.alfinito@unisalento.it, lino.reggiani@unisalento.it}
}

\maketitle

\begin{abstract}

We report on charge transport and current fluctuations in a single bacteriorhodpsin protein in a wide range of applied voltages covering direct and injection tunnelling regimes.
The satisfactory agreement between theory and available experiments validates the physical plausibility of the model developed here. 
In particular, we predict a rather abrupt increase of the  variance of current fluctuations in concomitance with that of the I-V characteristic.
The sharp increase, for about five orders of magnitude of current variance is associated with the opening of low resistance paths responsible for the sharp increase of the I-V characteristics.
A strong non-Gaussian behavior of the associated  probability distribution function is further detected by numerical calculations.

\end{abstract}

\IEEEpeerreviewmaketitle

\section{Introduction}
One of the more intriguing properties of proteins is the strong interdependence of  function and structure.  
Each protein consists of a specific sequence of aminoacids, spatially organized (folded) in few possible structures, each of them corresponding to a specific level of functioning. 
In the unfolded state the protein does not work. 
The switching from a structure to an other (say, for example, from the \textit{native} to the \textit{active} state) allows the protein to change its activity.
This feature has numerous technological applications, therefore it has recently received much attention from different fields of investigation. 
In particular, the structure switching can be used in the passive way to reveal the changing of functionality as induced by external agents. 
This is the case, for example, of recent investigations that point to the conformational change for revealing the capture of an external agent, like an odour \cite{Bond}. 
\par
The structure switching could  also be used in the active way, i.e. by inducing it with an external agent to modify some peculiar protein properties. 
An example of this application was obtained by a group at the Weizmann Institute \cite{Jin} which using a light sensitive protein, the bacteriorhodopsin (bR), was able to reveal a significant increase of the electrical current when the protein is irradiated with green light. 
This particular feature exploits the possibility to use the bR as the active element of an organic field effect transistor (OFET) \cite{ofet}, with the interesting traits of low cost, easy production and efficiency on a large range of voltages. 
In particular, the protein exhibits a good stability under the application of high fields, showing a response which explores the range from direct to injection  (Fowler-Nordheim) tunneling regime.
The interpretation and the theoretical modeling of this response was previously carried out by neglecting the protein internal structure \cite{Wang} and therefore it was not able to describe the modifications induced by the conformational change. 
\par
In this paper we propose a model of the electrical properties of proteins, based on the assumption that their modifications follow by their conformational change. 
The model has a wide applicability and allows the exploration of the topological properties of the proteins, such as of the static and dynamical responses under an electrical solicitations \cite{PRE}. 
Therefore, we are able both to describe  the existing data and to provide some  expectations. 
In particular, we reproduce the I-V characteristic given by experiments \cite{submitted} and also  predict  the associated fluctuations of the steady current.
Finally, by analyzing the  distributions of current fluctuations for different voltages, we find that they follow the universal behaviour
originally found by Bramwell-Holdsworth-Pintor (BHP) \cite{BHP}, and continuously rediscovered in most critical system \cite{BHP,SylosLabini,PhysicaA}
\section{Experiment and model}   
Bacteriorhodopsin is contained in a part of the cellular membrane, the purple membrane (PM), of the halophile \textit{Halobacterium salinarum} \cite{Corcelli}. 
PM is a particularly simple kind of cellular membrane, being constituted only by trimers of bR and some lipids to stabilize them, and is quite easy to produce. 
It appears as a quasi-2D structure, with a thickness of about 5 nm, corresponding to the protein dimension. 
These characteristics make it very suitable for any kind of experimental investigations.
\par
A seminal set of papers \cite{Jin} reported the measurements  carried on metal-insulator-metal (MIM) junctions of \textit{millimetric} diameter, with the insulator constituted by a 5 nm  monolayer of PM. 
The current response (nA level) was investigated in a small range of bias ($0 \div 1$ V), and in this range it appeared as  super-Ohmic, growing overall a factor of 2 when the sample is irradiated by a green light. 
These results suggest that in this protein, like in some
organic polymers \cite{Zvy}, electrical transport (ET) is ruled by tunneling mechanisms and that it is strongly related to the protein structure \cite{Epl}. 
In successive experiments \cite{Gomila}, the I-V characterization was performed at the \textit{nanometric} scale, with the conductive atomic force microscopy (c-AFM) technique.
In this technique, one of the contacts is constituted by the tip of the c-AFM
($100 \div 200$ nm of nominal radius). 
With respect to previous measurements, in the
common bias range, the current response is lower for about 4 order of magnitude, thus the sample is able to sustain higher
voltage up to about $5 \div 10$ V.
Measurements were performed without any extra light irradiation.
At voltages above about 2.5 V, the presence of a cross-over between
the direct tunneling regime and the injection or Fowler-Nordheim (FN) tunneling regime \cite{Simmons} was evidenced.
To our knowledge, a microscopic interpretation of the FN tunneling accounting for the structure of the protein is not present in literature, while some macroscopic approaches, which deal the protein as a uniform bulk insulator can be found \cite{Wang}. 
This paper aims to fill this gap by giving a tool for modeling the proteins at a microscopic level, and, at the same time, able to reproduce the macroscopic data. 
The model, hereafter called INPA (impedance network protein analogue), is based on a coarse-grained description of the protein: The aminoacids, taken as single centers of interaction, are the unique responsible of charge transfer and/or polarization. 
The protein is represented with an impedance network, which retains the main topological features of the protein \cite{Epl}, with a mechanism of electrical response that depends on the kind of applied bias (DC, AC) \cite{PRE}. 
In the present case, by using the analogy with the charge transport among localized states, we  model the  charge transport through a single protein by means of a \textit{sequential} tunneling mechanism \cite{Zvy}.
\par
The approach integrates structure and function of the protein and,
therefore, it constitutes a fundamental step toward the use of proteins like PM as an active part in bio-devices. 
With respect to previous approaches, INPA  has several advantages: 
$i)$  it can explain the different electrical responses at increasing voltages and in the presence/absence of green light;
$ii)$ it can reconcile the interpretation of different data \cite{Jin,Gomila}; 
$iii)$ it can be applied to other proteins whose tertiary structure is known.
\par
In brief the layout of the INPA is as follows.
The protein is mapped into an impedance network, by using the C$_\alpha$ atoms of each aminoacid as nodes. 
The number of links for each node is determined by the
cut-off interaction radius, say $R_{C}$.
In the present analysis, we choose $R_{C}$ = 6 \AA, a value that optimizes the native to activated state resolution \cite{Epl}. 
Each link is associated with an impedance (a simple resistance in this case)  whose value depends on the distance between amino-acids as:
$r_{i,j}=\rho\, l_{i,j}/\mathcal{A}_{i,j}$, 
where $\rho$ indicates the resistivity, here taken to be same for all the links, which in general depends on the voltage as detailed below, the pedices $i,j$ refer to the amino-acids between which the link is stretched,
$l_{i,j}$ is the distance between the labeled aminoacids taken as point like centers  and ${\mathcal{A}_{i,j}}$ is the cross-sectional area shared by the labeled amino-acids:
 ${\mathcal{A}_{i,j}}= \pi\left(\textsl{ R}_C^{2}-l^{2}_{i,j}/4\right)$.

To take into account the superlinear I-V response, the link resistivity, $\rho$,  is chosen to depend on the voltage drop of each resistance as:
\begin{equation}
\rho(V)=\left\{\begin{array}{lll}
\rho_{MAX}& \hspace{.5cm }& eV \leq  \Phi  \\ \\
 \rho_{MAX} (\frac{\Phi}{eV})+\rho_{min}(1- \frac{\Phi}{eV}) &\hspace{.5cm} & eV \ge  \Phi 
 \end{array}
  \right.
\label{eq:3}
\end{equation}
where $\rho_{MAX}$ is the resistivity value which should be used to fit the I-V characteristic at the lowest voltages,  $\rho_{min} \ll \rho_{MAX}$  plays the role of an extremely low series resistance, limiting the current at the highest voltages,  and $\Phi$  is the threshold energy separating the two tunneling regimes (a kind of effective height of a tunneling barrier).
Since ET is here interpreted in terms of a sequential tunneling between neighboring amino-acids, the above interpolation formula reflects the different voltage dependence in the prefactor of the current expression \cite{Wang}: $I\sim V$ in the direct tunneling regime, and $I\sim V^2$ in the FN tunneling regime.

\par
For  the transmission probability of the tunneling mechanism we take the expression given by Ref. \cite{Simmons}:
\begin{equation} 
\mathcal{P}^{D}_{i,j}= \exp \left[- \frac{2 l_{i,j}}{\hbar} \sqrt{2m(\Phi-\frac{1}{2}
eV_{i,j})} \right] \ ,
\hspace{0.7cm}
 eV_{i,j}  \leq \Phi  \,
\label{eq:1}
\end{equation}
\begin{equation}\label{eq:2}
\mathcal{P}^{FN}_{ij}=\exp \left[-\left(\frac{2l_{i,j}\sqrt{2m}}{\hbar}\right)\frac{\Phi}{eV_{i,j}}\sqrt{\frac{\Phi}{2}} \right] \ , 
\hspace{0.7cm}
 eV_{i,j} \ge \Phi  \ ;
\end{equation}
where $V_{i,j}$ is the local potential drop between the couple of $i,j$ amino-acids and $m$  is the electron effective mass, here taken the same of the bare value.
To model bacteriorhodopsin in its native state (in dark) \cite{Epl,PDB} we have taken the PDB entry 2NTU. 
The network is then studied with a point-like contacts configuration, connected to the external bias by means of perfectly conductive contacts, the input on the first amino-acid and the output on the last amino-acid of the primary structure. 
\section{Results}
\begin{figure}[htb]
	\centering
		\includegraphics[width=18pc]{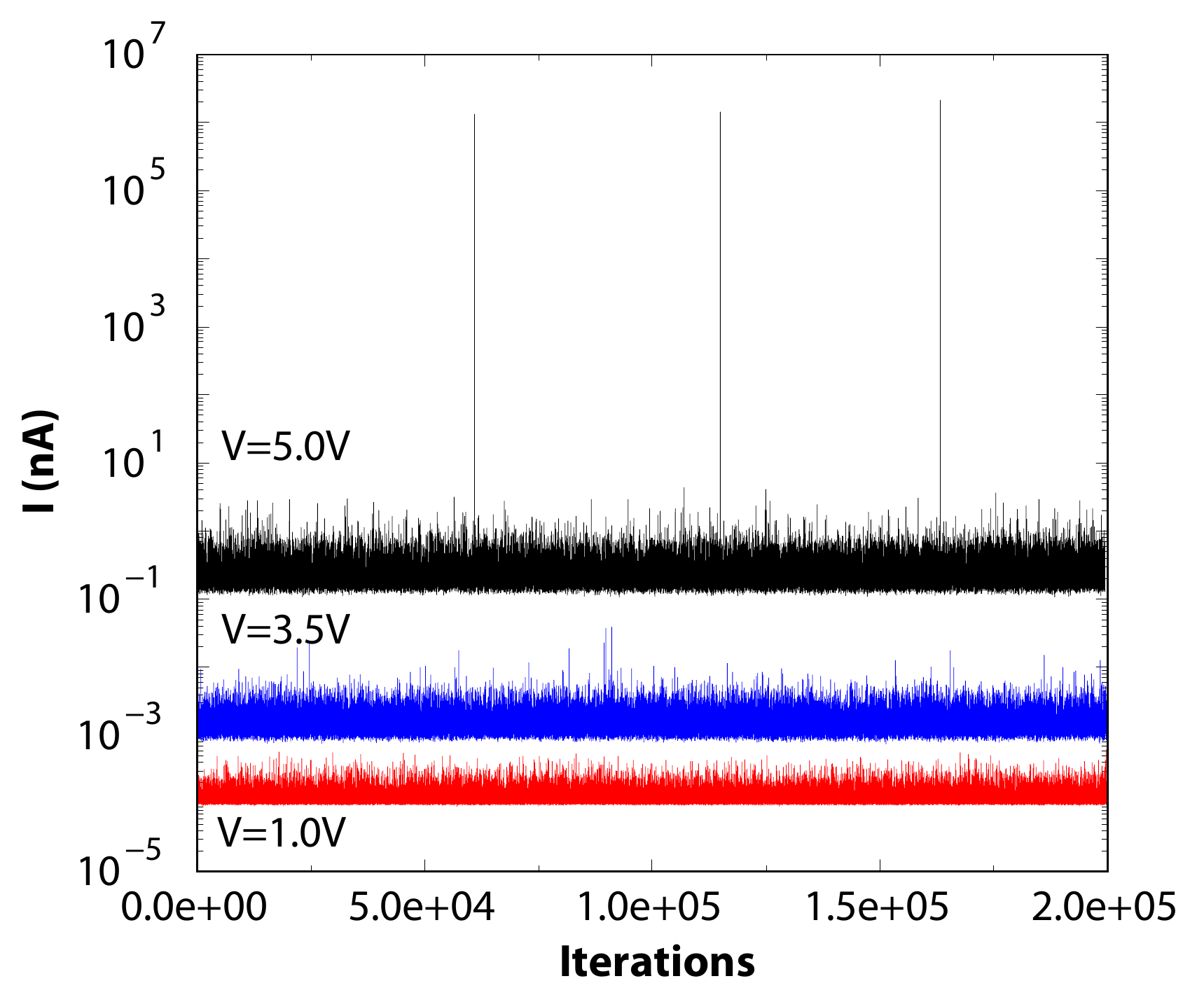}
		\caption{Current signal vs time (measured in number of iteration steps) at three different voltage values, from bottom to top: 1 V, 3.5 V, 5 V.}
		\label{fig:signal}
\end{figure}

Figure~(\ref{fig:signal}) reports the current evolution in bR as a function of
time (measured in number of iteration steps) for three different applied voltages: 1 V, in the low linear region; V=3.5 V in the transition region; V=5 V, approaching the extreme linear regime. 
The data relative to 5 V have been shifted up of 100 nA in order to better resolve the figure.
%
The current response exhibits a low level of noise at low voltages, where direct tunneling is the dominant transport process.
Near the transition between the two tunneling regimes, the amplitude of fluctuations increases, and finally, at high voltages an increasing number of very high current events (up to five order of magnitude larger than the major part of the signal) is manifested.
\par
\begin{figure}[htbp]
\centering\noindent
\includegraphics[width=18pc]{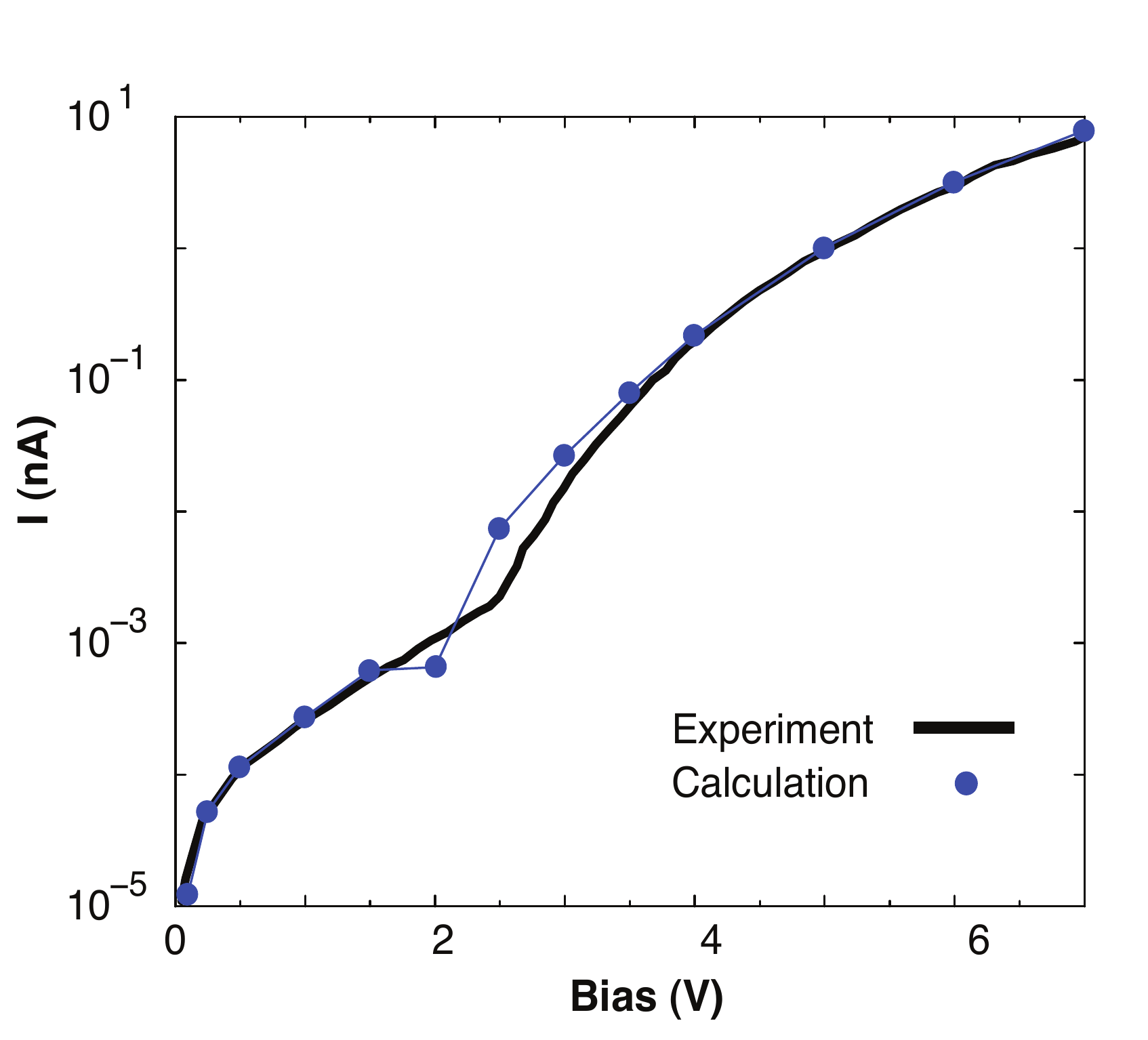}
\caption{I-V characteristic for bR at the nanometric scale: Experiments (bold line) \cite{Gomila} vs. simulations (circles), tiny line is a guide to eyes.
}
\label{fig:I-V}

\end{figure}

Figure~(\ref{fig:I-V}) compares the numerical and experimental data of the current voltage characteristic obtained with an AFM technique carried out on a bacteriorhodopsin monolayer at room temperature ~\cite{Gomila}. 
As relevant computational inputs, we have taken: 
$\Phi$, the threshold voltage value, of 219 meV, $\rho_{MAX} =4\cdot10^{13}\,
\Omega$ \AA , $\rho_{min}=4\cdot 10^{5}\,\Omega$ \AA . 
The satisfactory agreement between theory and experiments validates the physical plausibility of the model developed here. 
The sharp increase of the current above about 3V is associated with the cross-over between direct and injection sequential-tunneling regimes.
\par
\begin{figure}[htbp]
\centering\noindent
\includegraphics[width=18pc]{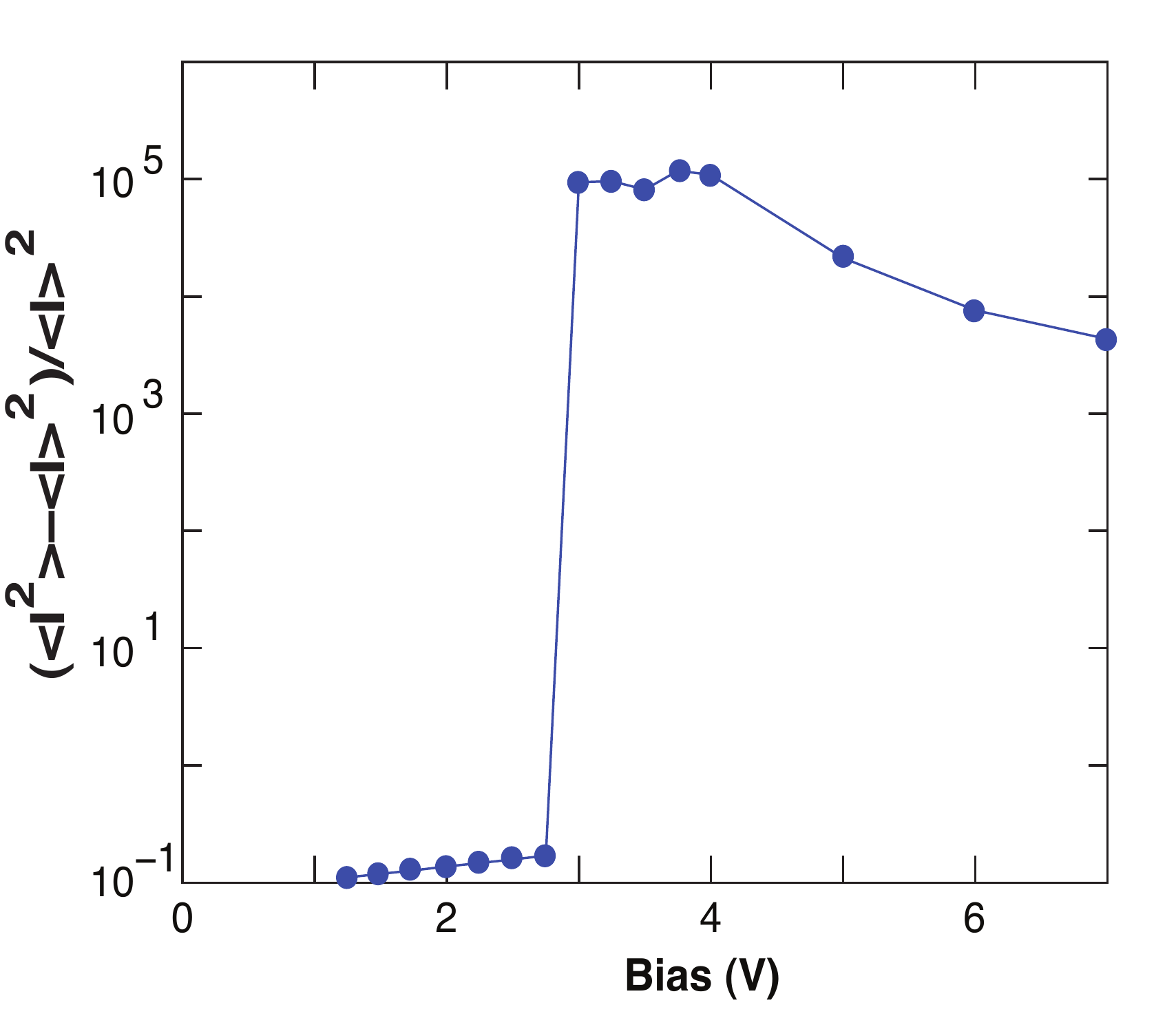}
\caption{Variance of current fluctuations obtained from simulations}
\label{fig:fluct}
\end{figure}

Figure~(\ref{fig:fluct}) reports the variance of current fluctuations corresponding to the I-V characteristic of  Fig. (\ref{fig:I-V}).
Here we notice a rather abrupt increase of the  variance of current fluctuations in concomitance with the cross-over region of the I-V characteristic.
The sharp increase for about five orders of magnitude of current variance is associated with the opening of low resistance paths responsible for the sharp increase of the I-V characteristics.
\noindent 
\noindent 
\begin{figure}[htb]
\centering\noindent
\includegraphics[width=18pc]{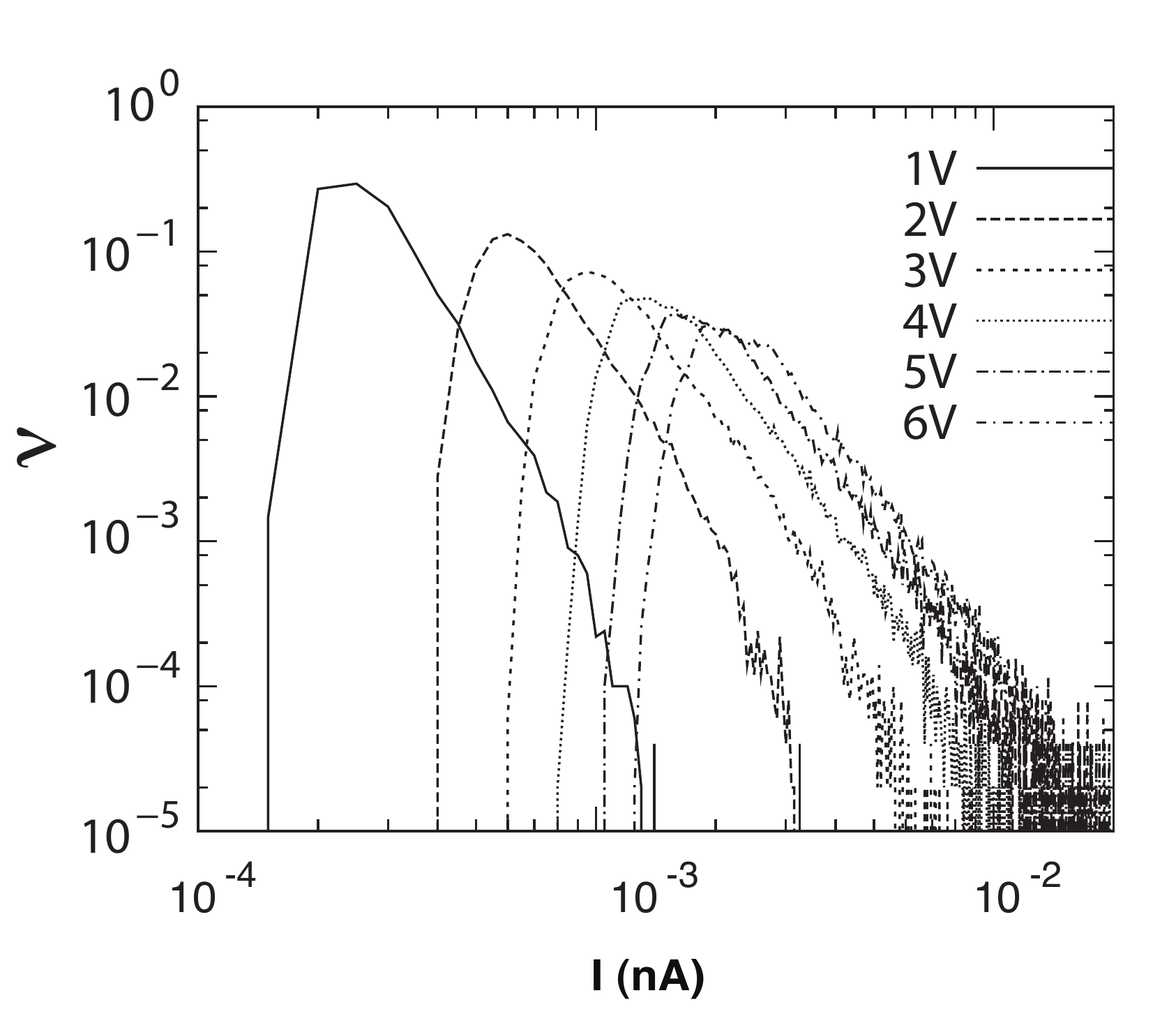}
\caption{Frequency distribution of currents for different values of applied voltages.}
\label{fig:gauss}
\end{figure}
By analising the frequency distribution of currents, $\nu$, for different voltages, a strong non-Gaussian behaviour can be observed (see Fig.(\ref{fig:gauss})). 
This result is not completely surprising, being partially due to the small size of the system under consideration \cite{PhysicaA} but also describing the critical condition of a system in which, for all the considered voltages, two different regimes are in competition. 
On the other hand, this kind of systems, and their fluctuations in particular, have received in the last few years \cite{BHP} a huge attention, mainly due to the possibility to reconduct them to a `universal' behaviour. 
This behaviour is described by a generalized Gumbel distribution, one of the reference distributions for the statistical modeling of extreme events \cite{Coles}. This distribution was extensively studied by Bramwell-Holdsworth-Pintor (BHP) who found it in very different contexts going from fluidynamics, to self-organized critical systems, and to resistance fluctuations.
The BHP distribution function has the following expression
\begin{equation} 
p(z)= A \,(e^{-z-e^{-z}})^a
\label{eq:bhp}
\end{equation}
where $p(z)$ is the probability density function (PDF).
We found that the skewed distributions reported in Fig.(\ref{fig:gauss}) can be simply recolapsed to a unique curve, by using as variable the natural logarithm of current and adopting the rescaled quantities:
\begin{equation}
\Pi(z)=\sigma \Phi(x), \qquad z=(x-<x>)/\sigma
\label{eq:rescale}
\end{equation}
with $\Phi(x),\ <x>,\ \sigma$ assuming the standard meaning of the probability density function, the mean value of the variable and the standard deviation, respectively.
The choice of the logarithm instead of the current is suggested by the slow dependence of this quantity on the applied voltage
\cite{BHP, SylosLabini}.
In particular, in a large voltage range, the distributions collapse and can be described by the expression: 
\begin{equation} 
p(z)=  a\,e^{-(ax+\gamma)-e^{-(ax+\gamma)}}
\label{eq:bhp}
\end{equation}
with $a=1.3$ and $\gamma$ the Euler constant \cite{SylosLabini}. 
\noindent 
\begin{figure}[htb]
\centering\noindent
\includegraphics[width=18pc]{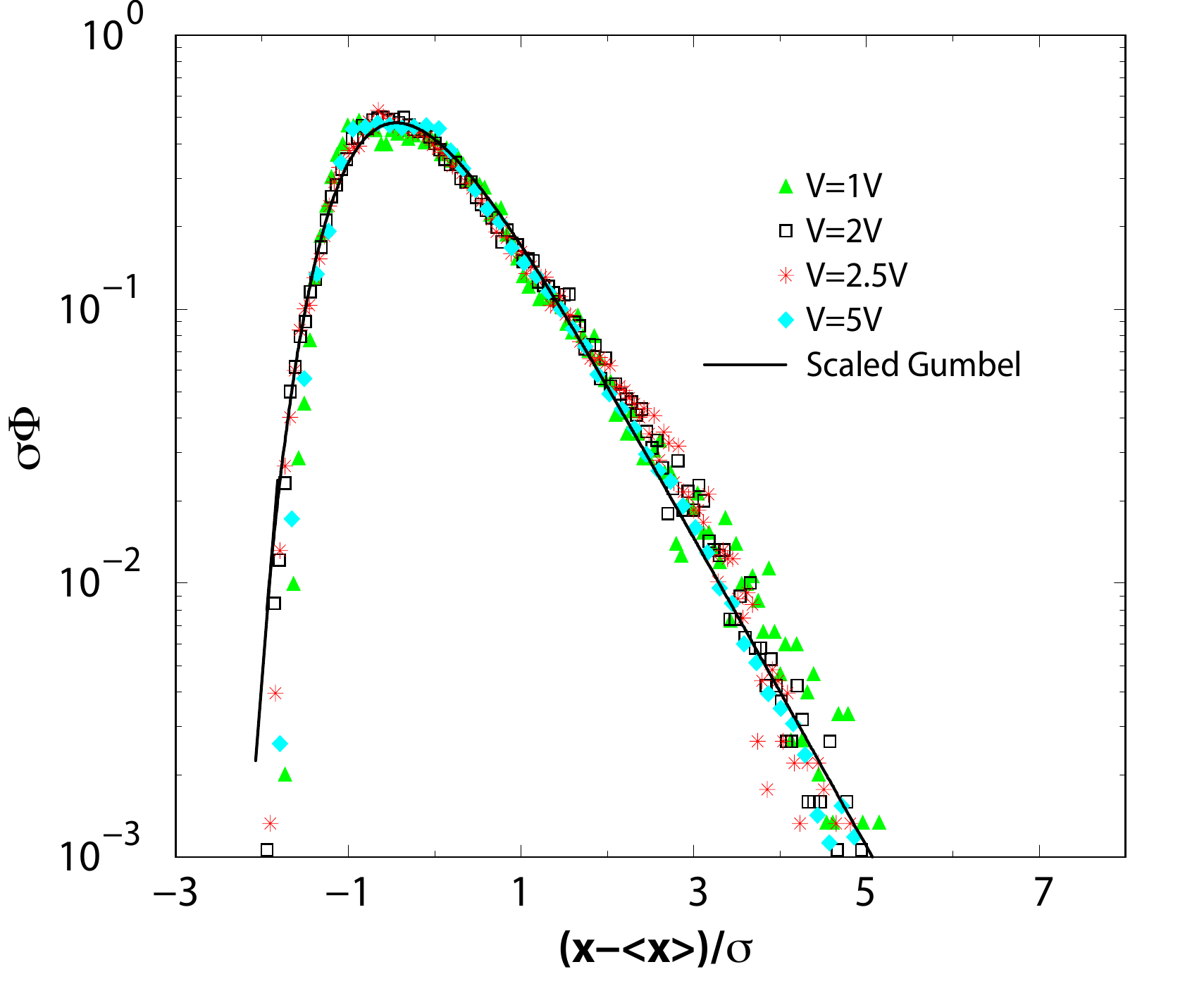}
\caption{Rescaled probability density function of current fluctuations for different values of applied voltage. The bold line is the scaled Gumbel distribution with a=1.3}
\label{fig:bhp}
\end{figure}
\section{Conclusion} 
The paper reports a microscopic model of current and fluctuations in proteins. 
In particular PM patches of the light receptor bacteriorhodopsin have been analyzed in a very wide range of voltage (0.01V - 7 V) where the current response is found to exhibit two different regimes of charge transport, direct and FN tunneling. 
A unique microscopic description, the so-called INPA model, is able to take into account both these regimes, thus producing the current response with a good agreement with available experimental data. 
Furthermore, the current fluctuations are found to follow a universal behaviour typical of  systems close to a phase transition. 
 \section*{Acknowledgment}
This research is carried out within the bioelectronic olfactory neuron device (BOND) project sponsored by the CE within the 7th Program, grant agreement: 228685-2.
\par
Dr G. Gomila is thanked for useful discussions on the subject.

\bibliographystyle{IEEEtran}
%


\begin{thebibliography}{1}
%

\bibitem{Bond}
Bioelectronic Olfactory Neuron Device (BOND), Collaborative project FP7-NMP-2008-SMALL-2, GA number 228685.
http://bondproject.org/
\bibitem{Jin} 
Y.~D.Jin, N.~Friedman, M.~Sheves, T.~He, and D.~Cahen, PNAS, \textbf{103}, 8601, 2006;
Y.~D. Jin, N.~Friedman, M.~Sheves, and D.~Cahen, Adv. Funct. Mater., \textbf{17}, 1417, 2007.
\bibitem{ofet}
 J. H. Sh{\"o}n, H. Meng, and Z. Bao, \emph{Self-assembled monolayer organic field-effect transistors}, Nature, 413, 713, 2001
%
\bibitem{Wang}
W. Wang, T. Lee, and W. A. Reed, Rep. Prog. Phys., \textbf{68}, 523, 2005.
%
\bibitem{PRE} E. Alfinito and L. Reggiani, Phys. Rev. E, \textbf{81}, 032902, 2010.
\bibitem{submitted} E. Alfinito, J.F. Millithaler and L. Reggiani, \emph{in preparation}
\bibitem{BHP}
S.~T.~Bramwell, P.~C.~W.~Holdsworth, and J.~-F. Pintor, \emph{Universality of rare fluctuations in turbulance and critical phenomena}, Nature, \textbf{396}, 444, 1998;S.~T.~Bramwell,\emph{The distribution of spatially averaged critical properties}, Nature Physics, \textbf{5},443 , 2009 (and references therein). 
\bibitem{SylosLabini}
T.~Antal, F.~ Sylos Labini, N.~ L.~ Vasilyev, and Y.~ B.~ Baryshev, \emph{Galaxy distribution and exteme-value statistics},
Eur.Phys. J. \textbf{88},59001 , 2009.
\bibitem{PhysicaA}
C.~Pennetta, E.~Alfinito, L.~Reggiani and S.~Ruffo, \emph{Non-Gaussian resistance noise near electrical breakdown in granular materials},  \textbf{340}, 380, 2004
%
\bibitem{Corcelli}
A. Corcelli, M. Coletta, G. Mascolo, F.P. Fanizzi, and M. Kates, \emph{A novel glycolipid and phospholipid in the purple membrane}, Biochemistry, \textbf{39}, 3318, 2000.

\bibitem{Zvy}
I. Zvyagin, \emph{Charge transport via delocalization in disordered materials} in S. Bartlett (Ed.), \emph{Charge transport in disordered materials} J.Wiley $\&$ Sons, The Atrium, Chichester, Southern Gate, West Sussex P019
8SQ, England 2008.


\bibitem{Epl} E.~Alfinito and L.~Reggiani, Europhys. Lett., \textbf{85}, 86002, 2009 and references therein.

\bibitem{Gomila}
I. Casuso, L. Fumagalli, J. Samitier, E. Padros, L. Reggiani, V. Akimov and G. Gomila, \emph{Electron transport through supported biomembranes at the nanoscale by conductive atomic force microscopy}, Nanotechnology, \textbf{18}, 465503, 2007.
%
\bibitem{Simmons}
R.H. Fowler, and L.W. Nordheim, Proc. R. Soc. London Ser. A, \textbf{119}, 173, 1928;
J.G. Simmons, \emph{Generalized formula for the electric tunnel effect between similar electrodes}, J. Appl. Phys., \textbf{34},1793, 1963.
%
\bibitem{PDB}H. M. Berman et al, Nucl.Acids Res., \textbf{28}, 235, 2000.
\bibitem{Coles}S.~ Coles, \emph{An introduction to statistical modeling of extreme values}, 3rd edition, London, Great Britain: Springer-Verlag, 2001.


%

%
%
%

%

 






%

 %
%
%





%
%


 %



\end{thebibliography}

\end{document}